\documentstyle[osa,manuscript]{revtex}

\begin{document}
\title{Unusual magnetic relaxation behavior in La$_{0.5}$Ca$_{0.5}$MnO$_{3}$ and Nd$%
_{0.5}$Sr$_{0.5}$MnO$_{3}$}
\author{J. L\'{o}pez*, P. N. Lisboa-Filho, W. A.\ C. Passos, W. A. Ortiz and F. M.
Araujo-Moreira}
\address{Grupo de Supercondutividade e Magnetismo, Departamento de F\'{i}sica,\\
Universidade Federal de S\~{a}o Carlos, Caixa Postal - 676, S\~{a}o Carlos,\\
SP, 13565-905, Brazil, *jlopez@df.ufscar.br}
\author{O. F. de Lima}
\address{Instituto de F\'{i}sica Gleb Wataghin, Universidade Estadual de Campinas,\\
UNICAMP, 13083-970, Campinas, SP, Brazil}
\author{D. Schaniel}
\address{Laboratory for Neutron Scattering, ETH Z\"{u}rich \& PSI Villigen, 5232\\
Villigen PSI, Switzerland}
\author{K. Ghosh}
\address{Physics,Astronomy and Materials Science Division,\\
Southwest Missouri State University, Springfield, MO 65804, USA}
\maketitle

\begin{abstract}
We have carried out a systematic magnetic relaxation study, measured after
applying and switching off a 5 T magnetic field to polycrystalline samples
of La$_{0.5}$Ca$_{0.5}$MnO$_{3}$ and Nd$_{0.5}$Sr$_{0.5}$MnO$_{3}$. The long
time logarithmic relaxation rate (LTLRR), decreased from 10 K to 150 K and
increased from 150 K to 195 K in La$_{0.5}$Ca$_{0.5}$MnO$_{3}$. This change
in behavior was found to be related to the complete suppression of the
antiferromagnetic phase above 150 K and in the presence of a 5 T magnetic
field. At 195 K, the magnetization first decreased, and after a few minutes
increased slowly as a function of time. Moreover, between 200 K and 245 K,
the magnetization increased throughout the measured time span. The change in
the slope of the curves, from negative to positive at about 200 K was found
to be related to the suppression of antiferromagnetic fluctuations in small
magnetic fields. A similar temperature dependence of the LTLRR was found for
the Nd$_{0.5}$Sr$_{0.5}$MnO$_{3}$ sample. However, the temperature where the
LTLRR reached the minimum in Nd$_{0.5}$Sr$_{0.5}$MnO$_{3}$ was lower than
that of La$_{0.5}$Ca$_{0.5}$MnO$_{3}$. This result agrees with the stronger
ferromagnetic interactions that exist in Nd$_{0.5}$Sr$_{0.5}$MnO$_{3}$ in
comparison to La$_{0.5}$Ca$_{0.5}$MnO$_{3}$. The above measurements
suggested that the general temperature dependence of the LTLRR and the
underlying physics were mainly independent of the particular charge ordering
system considered. All relaxation curves could be fitted using a logarithmic
law at long times. This slow relaxation was attributed to the coexistence of
ferromagnetic and antiferromagnetic interactions between Mn ions, which
produced a distribution of energy barriers.

\end{abstract}

\section{\protect\bigskip Introduction}

Besides the known magnetoresistance effect in manganese perovskites\cite
{Schiffer}, compounds like La$_{0.5}$Ca$_{0.5}$MnO$_{3}$\ and Nd$_{0.5}$Sr$%
_{0.5}$MnO$_{3}$\ present a real-space ordering of Mn$^{3+}$ and Mn$^{4+}$
ions, named as charge ordering (CO). These materials show, close to the
charge ordering temperature (T$_{CO}$), various anomalies in resistivity,
magnetization and lattice parameters as a function of temperature, magnetic
field and isotope mass\cite{Radaelli}$^{-}$\cite{Zhao2}. Microscopically, CO
compounds are particularly interesting due to the coexistence of
ferromagnetic and antiferromagnetic phases at low temperatures \cite
{Moritomo}. However, a relatively small external magnetic field destroys the
CO phase and enforces a ferromagnetic orientation of the spins \cite{Xiao}.
Moreover, electron microscope analysis has revealed convincing evidence that
the CO is accompanied by the orientational ordering of the 3d$^{3}$ orbitals
on the Mn$^{3+}$ ions\cite{Mori}, called as orbital ordering (OO). Recently,
X-ray resonant scattering experiments\cite{Zimmermann} suggested that CO (a
long range interaction) drove the OO (a short range interaction) near T$%
_{CO} $.

Neutron powder diffraction and magnetization studies in La$_{1-x}$Ca$_{x}$MnO%
$_{3}$ samples, with x=0.47, 0.50 and 0.53, have shown that the Curie
temperature (T$_{C}$) was approximately 265 K in all cases\cite{Huang}.
Huang et. al. \cite{Huang} also reported the formation of a second
crystallographic phase (A-II) at 230 K. Moreover, they found that the A-II
phase had the same space-group symmetry (Pnma) and lattice parameters as the
original F-I phase, but it differed in the weak Jahn-Teller distortions of
the MnO$_{6}$ octahedrons. Furthermore, the A-II phase ordered
antiferromagnetically with a CE-type magnetic structure below 160 K.
Besides, both Huang et. al. \cite{Huang} and Radaelli et. al. \cite{Radaelli}
observed a rapid change of the lattice parameters between 130 and 225 K in La%
$_{0.5}$Ca$_{0.5}$MnO$_{3}$. This was associated with the development of a
Jahn-Teller distortion of the Mn-O octahedra, as well as partial orbital
ordering.

The physical properties in CO manganese perovskites arise from the strong
competition among a ferromagnetic double exchange interaction, an
antiferromagnetic superexchange interaction, and the spin-phonon coupling.
These interactions are determined by intrinsic parameters such as doping
level, average cationic size, cationic disorder and oxygen stoichiometry.
Magnetic relaxation studies are a useful tool to study the dynamics of these
competing interactions.

Fisher et. al. \cite{Fisher} studied the time relaxation of resistivity and
magnetization in colossal magnetoresistance compounds like La$_{1-x}$A$_{x}$%
MnO$_{3}$ (with A=Sr, Ca and x=0.3, 0.35). They found that the relaxation
rate was temperature dependent and slowed down when the temperature was far
from T$_{C}$. The spin-spin relaxation time ($\tau _{ss}$) as a function of
temperature in La$_{0.5}$Ca$_{0.5}$MnO$_{3}$ was measured by Dho et. al. 
\cite{Dho} using $^{55}$Mn and $^{139}$La nuclear magnetic resonance. They
found a general decrease in $\tau _{ss}$ with increasing temperatures.
However, $\tau _{ss}$ showed hysteretic behavior in zero field cooling and
field cooling measurements, approximately in the same temperature interval
where the change in lattice parameters was most pronounced.

The relaxation of electrical resistivity in Pr$_{0.67}$Ca$_{0.33}$MnO$_{3}$,
after a large change in applied magnetic field, which induces a transition
from a ferromagnetic metallic state to a charge ordered insulator phase or
conversely, was studied by Anane et. al. \cite{Anane}. They found an abrupt
change in the resistivity, which indicated a metal to insulator transition.
Smolyaninova et. al. \cite{Smolyaninova} reported the time dependence of the
resistivity and magnetization in a La$_{0.5}$Ca$_{0.5}$MnO$_{3}$ sample at
low temperatures. They fitted all the curves to a stretched exponential time
dependence and explained their results using a hierarchical distribution of
relaxation times.

However, to our knowledge, studies of magnetic relaxation curves (M(t)) in
charge ordered compounds are not reported for a wide temperature interval.
Here, we present a general magnetic characterization of two polycrystalline
samples of La$_{0.5}$Ca$_{0.5}$MnO$_{3}$ and Nd$_{0.5}$Sr$_{0.5}$MnO$_{3}$.
Besides, in both cases M(t) curves were systematically measured for several
temperatures. We found an unusual temperature dependence of the relaxation
curves in the region approximately between the Neel (T$_{N}$) and Curie
temperatures. These results could be interpreted as a consequence of the
strong competition between ferromagnetic and antiferromagnetic interactions.
A short version of these results were reported elsewhere\cite{yo-icm2000} by
some of us.

\section{Experimental procedures}

Polycrystalline samples of La$_{0.5}$Ca$_{0.5}$MnO$_{3}$ were prepared from
stoichiometric amounts of La$_{2}$O$_{3}$, CaCo$_{3}$, and Mn$_{3}$O$_{4}$
by standard solid-state reaction method. Purity of these starting materials
was more than 99.99 \%. As most of the rare earth oxides absorb moisture
from the air, La$_{2}$O$_{3}$ has been preheated at 1000 $^{o}$C for 12
hours. All the powders were mixed and grinded for a long time in order to
produce a homogeneous mixture. First, the mixture was heated at 1100 $^{o}$C
for 12 hours and after that it was grinded and heated several times at 1100 $%
^{o}$C, 1200 $^{o}$C and 1300 $^{o}$C. After the single-phase material was
reached, as checked by X-ray scattering, a pellet was pressed and sintered
at 1400 $^{o}$C for 24 hours.

Polycrystalline samples of Nd$_{0.5}$Sr$_{0.5}$MnO$_{3}$ were prepared by
the sol-gel method \cite{Paulo}. Stoichiometric parts of Nd$_{2}$O$_{3}$ and
MnCO$_{3}$ were dissolved in HNO$_{3}$ and mixed to an aqueous citric acid
solution, to which SrCO$_{3}$ was added. The mixed metallic citrate solution
presented the ratio citric acid/metal of 1/3 (in molar basis). Ethylene
glycol was added to this solution, to obtain a citric acid/ethylene glycol
ratio 60/40 (mass ratio). The resulting blue solution was neutralized to pH$%
\sim $7 with ethylenediamine. This solution was turned into a gel, and
subsequently decomposed to a solid by heating at 400 $^{o}$C. The resulting
powder was heat-treated in vacuum at 900 $^{o}$C for 24 hours, with several
intermediary grindings, in order to prevent formation of impurity phases.
This powder was pressed into pellets and sintered in air at 1050 $^{o}$C for
12 hours. X-ray diffraction measurements did not show any peak associated to
either impurities or starting materials, indicating the high quality of the
samples.

The magnetization measurements reported here were done with a Quantum Design
MPMS-5S SQUID magnetometer (UFSCar-S\~{a}o Carlos). Besides, in order to
rule out possible machine-dependent effects we repeated part of the
relaxation measurements using different techniques and equipments (a Quantum
Design PPMS magnetometer operating with the extraction method and another
Quantum Design MPMS SQUID magnetometer) in two other laboratories
(ETH-Zurich and Unicamp-Campinas, respectively). The results were consistent
and reproducible in all cases.

The relaxation measuring procedure was the following: first, the sample was
heated to 400 K in zero magnetic field; second, the remanent magnetic field
in the solenoid of the SQUID magnetometer was set to zero; third, the sample
was cooled down in zero magnetic field until the stable working temperature
was reached; fourth, an applied magnetic field (H) was increased from 0 to 5
T at a rate of 0.83 T/minute and maintained for a waiting time t$_{w}$=50 s;
fifth, H was decreased to the end field (H$_{end}$) of zero at the same
rate; finally, when H was zero (we defined this time as t=0) the M(t) curve
was recorded for approximately 3 hours. In the future we will call this
procedure as ``the standard''. However, in the case of figure 5,
measurements were done after step four. We have measured also the profile of
the remanent magnetic field trapped in the superconducting solenoid after
increasing H to 5 T and its subsequent removal. Within the experimental
region the trapped magnetic field was smaller than 1.1 mT.

\section{\protect\bigskip Results and Discussion}

\subsection{Magnetization versus temperature measurements}

Figures 1a and 1b show the temperature dependence of the magnetization
measured with H=5 T for La$_{0.5}$Ca$_{0.5}$MnO$_{3}$ and Nd$_{0.5}$Sr$%
_{0.5} $MnO$_{3}$ samples, respectively. The insets show the same type of
measurements with (a) H=1.2 mT and (b) H=0.1 mT. Data in the main frame are
shown for zero field cooling (ZFC), field cooling (FC) and field cooling
warming (FCW) conditions (see the arrows), while the ones in the insets are
shown for FC and FCW conditions. The large hysteresis at a magnetic field as
high as 5 T is a clear evidence of the intrinsic frustration in the
equilibrium configuration of the spin system, which could not be only
associated to grain boundary defects. This feature makes these compounds
particularly interesting to study their relaxation behavior.

We fitted the small field magnetization data for both samples, using
temperatures above 300 K, to a Curie-Weiss law: $M/H\sim \mu
_{eff}^{2}/(T-T_{\Theta })$, where $M/H$ was the DC susceptibility, $\mu
_{eff}$\ was the effective paramagnetic moment and $T_{\Theta }$\ was the
Curie-Weiss temperature. The $\mu _{eff}$ and $T_{\Theta }$ values were 5.5 $%
\mu _{B}$ and 5.7 $\mu _{B}$, and 252 K and 254 K for La$_{0.5}$Ca$_{0.5}$MnO%
$_{3}$ and Nd$_{0.5}$Sr$_{0.5}$MnO$_{3}$ samples, respectively. The positive
values of the Curie-Weiss temperatures were an indication of the
preferential ferromagnetic interaction between spins in this temperature
range.

Effective paramagnetic moments could be compared to a simplified theoretical
model\cite{Millange}. Taking the orbital momentum to be quenched in both Mn$%
^{3+}$ and Mn$^{4+}$, the theoretical effective paramagnetic moment ($\mu
_{eff}^{Th}$) in each case reduces to the spin contribution ${\em g\ }\sqrt{%
{\em S\ (S+1)}}{\em \ }\mu _{B}$, where {\em S} is the spin of the ion (3/2
for Mn$^{4+}$ and 2 for Mn$^{3+}$) and {\em g} is the gyromagnetic factor
(approximately 2 in both cases). That gives the theoretical values of $\mu
_{eff}^{Th}(Mn^{3+})=4.90$\ $\mu _{B}$ and $\mu _{eff}^{Th}(Mn^{4+})=3.87$\ $%
\mu _{B}$. Then, assuming 0.5 Mn$^{3+}$ and 0.5 Mn$^{4+}$ ions per formula
unit, it is found $\mu _{eff}^{Th}(La_{0.5}Ca_{0.5}MnO_{3})=\sqrt{\left[ 0.5%
\left[ \mu _{eff}^{Th}(Mn^{3+})\right] ^{2}+0.5\left[ \mu
_{eff}^{Th}(Mn^{4+})\right] ^{2}\right] }=4.41$\ $\mu _{B}$.

In the case of the Nd$_{0.5}$Sr$_{0.5}$MnO$_{3}$ sample the magnetic moment
of the Nd$^{3+}$ ions should also be included. The electronic levels of Nd$%
^{3+}$ ions at high temperatures are well described by {\em g=8/11} and the
total angular momentum {\em J=9/2}, which leads to $\mu _{eff}^{Th}(Nd^{3+})=%
{\em g}\sqrt{{\em J(J+1)}}\mu _{B}=3.62$\ $\mu _{B}$. Therefore, considering
a rigid coupling of the moments of Nd$^{3+}$ ions with the moments of Mn$%
^{3+}$ and Mn$^{4+}$ ions, we should have\ $\mu
_{eff}^{Th}(Nd_{0.5}Sr_{0.5}MnO_{3})=\sqrt{\left[ 0.5\left[ \mu
_{eff}^{Th}(Mn^{3+})\right] ^{2}+0.5\left[ \mu _{eff}^{Th}(Mn^{4+})\right]
^{2}+0.5\left[ \mu _{eff}^{Th}(Nd^{3+})\right] ^{2}\right] }=5.10$\ $\mu _{B}
$. The experimental values of the effective paramagnetic moments are higher
than the theoretical ones in both samples. This could be a consequence of
cluster formation of Mn$^{4+}$ and Mn$^{3+}$ ions at high temperatures.
Similar high values of the paramagnetic effective moments in samples of La$%
_{1-x}$MnO$_{3}$ were reported by S. de Brion et. al.\cite{Brion}.

The derivative of the curves in the insets of figure 1 showed minima values
around 230 K and 250 K for La$_{0.5}$Ca$_{0.5}$MnO$_{3}$ and Nd$_{0.5}$Sr$%
_{0.5}$MnO$_{3}$ samples, respectively. This criterion has been usually
employed as a definition of T$_{C}$. The maximum magnetization for the FCW
curves, using the smaller applied field, was found at 210 K for La$_{0.5}$Ca$%
_{0.5}$MnO$_{3}$ and at 176 K for Nd$_{0.5}$Sr$_{0.5}$MnO$_{3}$. These peaks
indicated that antiferromagnetic correlations already existed at these
temperatures. However, the Neel temperature, found from neutron diffraction
studies, was T$_{N}$=160 K in both cases \cite{Radaelli}$^{,}$ \cite{Huang}$%
^{,}$ \cite{Kajimoto}$^{,}$ \cite{Kawano}.

The peak positions in the magnetization curves for both samples are strongly
dependent on the cooling conditions and the applied magnetic field. This
emphasizes the small energy differences among distinct equilibrium spin
configurations. Magnetization maxima occur at lower temperatures in Nd$%
_{0.5} $Sr$_{0.5}$MnO$_{3}$ than in La$_{0.5}$Ca$_{0.5}$MnO$_{3}$. Besides,
differences between FC and FCW curves, and correspondingly, the hysteresis
loop area, are smaller for Nd$_{0.5}$Sr$_{0.5}$MnO$_{3}$. These results
correlate with the stronger ferromagnetic interactions in Nd$_{0.5}$Sr$%
_{0.5} $MnO$_{3}$, which is evidenced by its higher T$_{C}$ and a higher
value of the magnetization at 2 K.

Ferromagnetic and antiferromagnetic phases coexist at low temperatures for
both compounds. The spins at low temperatures align in a
CE-antiferromagnetic lattice \cite{Kajimoto}$^{,}$ \cite{Kawano}, which is
also charge and orbital ordered. As we saw before, in contrast to the La$%
^{3+}$ ions, the Nd$^{3+}$ ions have an intrinsic total angular momentum
(J=9/2). Figure 1b shows an increase in the magnetization at temperatures
approximately below 50 K, indicating a possible short-range magnetic order
of the Nd$^{3+}$ ions. A similar increase in the magnetization at low
temperatures has been reported for Nd$_{0.5}$Ca$_{0.5}$MnO$_{3}$\cite
{Millange}. Besides, a low temperature specific heat study in Nd$_{0.67}$Sr$%
_{0.33}$MnO$_{3}$\cite{Gordon}, found a Schottky-like peak correlated to the
ordering of the Nd$^{3+}$ ions. Our recent specific heat measurements in Nd$%
_{0.5}$Sr$_{0.5}$MnO$_{3}$ have also confirmed these results.

\subsection{\protect\bigskip Magnetization versus field measurements}

Figure 2 shows representative magnetization versus field curves (M vs. H)
for the La$_{0.5}$Ca$_{0.5}$MnO$_{3}$ (2a and 2b) and Nd$_{0.5}$Sr$_{0.5}$MnO%
$_{3}$ (2c and 2d) samples, respectively. The applied magnetic field was
cycled in all cases from 0 to 5 T and then back to 0 T again. At 350 K both
samples are in the paramagnetic state and the M vs. H curves are linear. The
first indication of non-linear behavior is seen at small fields around 270
K. Below 230 K a rapid increase in magnetization at small field values is
well defined. This last behavior will be present for all lower temperatures,
identifying an easily oriented ferromagnetic component. It is already noted
in the curves for 230 K that at small fields the slope has a larger value
for the Nd$_{0.5}$Sr$_{0.5}$MnO$_{3}$ sample. This is in agreement with the
stronger ferromagnetic interactions in the Nd$_{0.5}$Sr$_{0.5}$MnO$_{3}$
sample.

The start of the hysteretic behavior in the M vs. H curves, at about 190 K
for La$_{0.5}$Ca$_{0.5}$MnO$_{3}$\ and 160 K for Nd$_{0.5}$Sr$_{0.5}$MnO$%
_{3} $, correlates with the separation of FC and FCW condition curves in
figures 1a and 1b. The area of the hysteresis loop first increases for lower
temperatures and then decreases again. The hysteresis loop area is largest
at 160 and 140 K for La$_{0.5}$Ca$_{0.5}$MnO$_{3}$\ and Nd$_{0.5}$Sr$_{0.5}$%
MnO$_{3}$ samples, respectively. Besides, the remanent magnetization, after
increasing the applied magnetic field to 5 T and decreasing it again to
zero, is maximum at 180 K for La$_{0.5}$Ca$_{0.5}$MnO$_{3}$, corresponding
to only 2.7 \% of the magnetization value at 5 T. However, for the Nd$_{0.5}$%
Sr$_{0.5}$MnO$_{3}$\ case, the remanent magnetization is maximum at 160 K,
corresponding to 13 \% of its value at 5 T.

At about 170 K and fields around 1 T, a linear magnetization field
dependence with a small slope, characteristic of the gradual destruction of
an antiferromagnetic phase, starts to be observed in both samples. This
small-slope linear behavior disappears around 3 T, because a ferromagnetic
phase is induced by the high magnetic field. The complete suppression of the
antiferromagnetic phase is no longer seen in our data below about 150 K,
because magnetic fields higher than 5 T would be required. Gang Xiao et. al. 
\cite{Xiao} reported M vs. H curves for La$_{0.5}$Ca$_{0.5}$MnO$_{3}$ with
fields up to 20 T. They found that the field for a complete destruction of
the antiferromagnetic phase increased from approximately 3.5 T at 164 K to
11 T at 4.2 K and that the transition was of first order. Similar results
have been reported for Nd$_{0.5}$Sr$_{0.5}$MnO$_{3}$\cite{Shimomura}.

In a simplified thermodynamical model, the ferromagnetic (FM) and
antiferromagnetic (AFM)\ charge ordered states correspond to two local
minima in the free energy, with large and small magnetization values,
respectively. A potential barrier {\em U} separates these local minima. An
external magnetic field tends to stabilize the FM state due to the gain (%
{\em - M }$\cdot ${\em \ H}) in the free energy rather than the AFM charge
ordered state ({\em M}$\sim ${\em 0}). Hence, the potential barrier is
expected to vary with the field. The sum of the Zeeman and thermal energies
should cover the energy difference between the AFM and FM states in order to
favor the AFM-FM transition, explaining the smaller transition magnetic
fields for higher temperatures.

Figure 2c further shows that for high fields the magnetization at 2 K
(dotted lines) is higher than at 20 K (open down triangles). This result
correlates with the increase in magnetization at low temperatures, observed\
for Nd$_{0.5}$Sr$_{0.5}$MnO$_{3}$ (fig.1b) but not for La$_{0.5}$Ca$_{0.5}$%
MnO$_{3}$ (fig.1a). As we have mentioned before, Nd$^{3+}$ ions, unlike La$%
^{3+}$ ions, have an intrinsic magnetic moment. Thus, this increase in
magnetization could be related to short range magnetic ordering of the Nd$%
^{3+}$ ions at low temperatures \cite{Gordon}. Above 1.5 K no long range
order of Nd$^{3+}$ ions was detected in neutron diffraction studies of Nd$%
_{0.5}$Ca$_{0.5}$MnO$_{3}$\cite{Millange}.

\subsection{Relaxation measurements}

Figure 3 shows magnetic relaxation measurements, after applying and removing
a 5 T magnetic field, from (a) 10 to 150 K, (b) 150 to 195 K and (c) 195 to
245 K in a La$_{0.5}$Ca$_{0.5}$MnO$_{3}$ sample. To facilitate the
comparison between curves at different temperatures, the magnetization in
each case was normalized to the corresponding value at {\em t=0}, and time
was plotted in a logarithmic scale. These curves were denoted as {\em %
m(t)=M(t)/M(0)}. The mean slope of each curve at long times, or long time
logarithmic relaxation rate (LTLRR), decreases systematically with
increasing temperatures from 10 to 150 K. It is important to note that
slopes in figures 3a and 3b are negative. The fractional change in
magnetization, between the first and the last measurement, increases from
1\% at 10 K to about 20 \% at 150 K. This qualitative behavior has been
usually explained considering the increase in the thermal energy, which
stimulates the random alignment of the spin and, as a consequence, a
decrease in the magnetization.

As can be seen in figure 3b, and contrary to the previous interval, the
LTLRR between 150 and 195 K increases with temperature. The fractional
change in magnetization goes from 20 \% at 150 K to 0.9 \% at 195 K. Note
from figure 2a, that above 150 K an applied magnetic field of 5 T completely
destroys the antiferromagnetic phase. These results correlate with the
change in the temperature dependence of the LTLRR at 150 K. Roughly
speaking, between 150 and 195 K the system seems to remember its previous 5
T ferromagnetic orientation. After removing the applied magnetic field, it
starts to relax at a slower rate, even in the presence of greater thermal
excitations. We will see below that, in this temperature range, the increase
in thermal energy is compensated by a rapid increase in the pinning energy
of the spin system.

The inset in figure 3b reproduces the relaxation curve before normalization
for 195 K and the corresponding error bars. Notice that, in spite of the
experimental error, this curve clearly shows that the magnetization first
decreases, and after approximately 4 minutes, increases with time. This
behavior is qualitatively different from the curves measured at lower
temperatures. We would like to stress here that this unusual increase in the
magnetization is stable during a long time interval (about 3 hours). The
inset in figure 3c shows the M(t) curve for 200 K, error bars here are about
the same size of the symbols. In this case, no decrease in magnetization was
measured, but a monotonic increase with time (approximately 62 $\cdot $ 10$%
^{-5}$ $\mu _{B}$ per Mn ion in 218 minutes) was observed. The experiment at
200 K was also repeated with a sample of the same compound having only 9 \%
of the original mass. A similar increase of magnetization with time (88 $%
\cdot $ 10$^{-5}$ $\mu _{B}$ per Mn ion in 218 minutes) was found in this
last case.

Figure 3c shows {\em m(t)} curves, using the standard relaxation procedure,
from 195 to 245 K. Notice that in contrast to figure 3b, all curves in
figure 3c, except the one at 195 K, show values above one. In other words,
the magnetization increases with time (curves here have positive slopes)
above the {\em M(0)} value in each case. Furthermore, the fractional change
in magnetization is systematically higher with higher temperatures: from 0.9
\% at 195 K to 80 \% at 240 K. However, {\em M(0)} decreases with higher
temperatures, as shown in figure 1a.

We also repeated the standard relaxation procedure at 210 K, but now with an
increasing waiting time in each case: {\em t}$_{w}${\em =50 s}, {\em t}$_{w}$%
{\em =500 s} and {\em t}$_{w}${\em =5000 s}. We would like to stress that in
the last case {\em t}$_{w}$ was longer than one hour. The normalized
increment in the magnetization was higher the longer the 5 T magnetic field
remained applied. Values of {\em M(0)} also increased for longer {\em t}$_{w}
$. These measurements confirmed the presence of the unusual relaxation,
independently of the value of the waiting time. A plausible explanation
could be that the remanent trapped field in the sample after removing the
H=5 T, which was higher for longer {\em t}$_{w}$, was causing a
self-alignment of the spins and an increase in magnetization. Therefore,
these results could be reflecting intrinsic information about the
interactions in the sample.

The curve at 245 K, also shown in figure 3c, presents a smaller fractional
change in magnetization with respect to the one at 240 K. This is probably
associated with the transition of the system to the paramagnetic phase.
Magnetic relaxation measurements were also done between 245 K and 350 K. In
this temperature interval we did not find a systematic variation of the
LTLRR, probably due to the small values of the absolute magnetization.
Nonetheless, above 260 K, the {\em M(t)} curves always showed the usual
decreasing behavior with time.

The change in the temperature dependence of the LTLRR above 195 K could be
correlated with the gradual disappearance of the antiferromagnetic
fluctuations even at small applied magnetic fields. This is reflected in the
gradual suppression of the hysteresis in the M vs. H curves (see figure 2b),
and also by the peak in the FCW curve in the inset of figure 1a. However,
although antiferromagnetic fluctuations are reduced for higher temperatures,
the ferromagnetic interactions also weaken. Therefore, at a given
temperature, the competition of these two effects causes the system to
return to the usual relaxation behavior.

It is also interesting to note the close overlap between the whole
temperature interval where the unusual magnetic relaxation was found (150 K
to 245 K) and the temperature interval where a rapid change in the lattice
parameters (130 K to 230 K) were reported\cite{Huang}$^{,}$\cite{Radaelli}.
This rapid change has been associated to the development of a Jahn-Teller
distortion of the Mn-O octahedra, as well as to the partial orbital ordering
of Mn ions\cite{Huang}$^{,}$\cite{Radaelli}. Therefore, the close overlap
between both temperature intervals suggests that the electron-phonon
interaction needs to be considered to completely understand this unusual
relaxation.

Recently, similarly unusual magnetic relaxation measurements were done by
Sirena et. al.\cite{Sirena}. They studied the relaxation of the
magnetization in thin films of La$_{0.6}$Sr$_{0.4}$MnO$_{3}$, after applying
a 1 T magnetic field during 5 minutes and then removing it. Their
measurements were done during a time window of 8 hours. This procedure was
repeated for temperatures between 4 K and 200 K. They found that, above a
temperature labeled as {\em T}$_{rev}$, the magnetization increased with
time and {\em T}$_{rev}$ decreased with increasing film thickness. However,
no clear interpretation was reported for all the results.

Figure 4 shows magnetic relaxation measurements using the standard procedure
in the Nd$_{0.5}$Sr$_{0.5}$MnO$_{3}$ sample. The LTLRR was negative at low
temperatures and decreased from 10 to 130 K (fig. 4a). From 130 to 170 K
(fig. 4b) the LTLRR was still negative but increased as a function of
temperature. The LTLRR became positive, increasing even more, from 180 to
250 K (fig. 4c). The absolute variation in magnetization, between the first
and the last measurement, was 1 \% at 10 K, 3 \% at 130 K and 5 \% at 250 K.
We also performed magnetic relaxation measurements above 250 K, but we did
not find a systematic variation of the LTLRR, maybe due to the small values
of the absolute magnetization in this temperature range. The temperature
where the LTLRR reached the minimum in Nd$_{0.5}$Sr$_{0.5}$MnO$_{3}$ was
lower than in La$_{0.5}$Ca$_{0.5}$MnO$_{3}$. This is in agreement with the
lower temperature of the maximum FC-magnetization for Nd$_{0.5}$Sr$_{0.5}$MnO%
$_{3}$\ (see figure 1).

The relaxation curves measured for Nd$_{0.5}$Sr$_{0.5}$MnO$_{3}$\ followed
the qualitative behavior found for La$_{0.5}$Ca$_{0.5}$MnO$_{3}$. Note also
in figure 2c that, above 130 K and closely related to the change in behavior
of the LTLRR, a 5 T applied magnetic field completely destroys the
antiferromagnetic phase. Besides, as in the La$_{0.5}$Ca$_{0.5}$MnO$_{3}$
case, there is a rapid change in the lattice parameters for Nd$_{0.5}$Sr$%
_{0.5}$MnO$_{3}$\ between approximately 110 K and 250 K. Once more, both
temperature intervals (rapid change of lattice parameters and unusual
magnetic relaxation behavior) almost completely overlap. All of these
suggest that the temperature dependence of the LTLRR, and the underlying
physics, are mainly independent of the particular charge-ordering material.

Moreover, in order to test the effects of a fixed applied magnetic field on
the temperature dependence of the LTLRR, we repeated the standard relaxation
procedure in the La$_{0.5}$Ca$_{0.5}$MnO$_{3}$ sample, but in this case we
did not remove the 5 T field. Figure 5 shows these relaxation curves and a
schematic drawing of the time evolution of the applied magnetic field. As
before, the magnetization was normalized and time was shown in logarithmic
scale. Here, the LTLRR is positive and increases from 10 to 150 K (fig. 5a)
and decreases from 150 to 170 K (fig. 5b). From 190 to 210 K (fig. 5c) the
LTLRR is negative and decreases with increasing temperatures.

The absolute variation in magnetization between the first and the last
measurement is 1.2 \% at 10 K, 3.4 \% at 150 K and 0.6 \% at 210 K. These
variations are smaller than in the previous cases, due to the high value of
the applied magnetic field. The maximum LTLRR is found here at 150 K, the
same temperature where we had found the minimum LTLRR for the case with {\em %
H}$_{end}${\em =0 T} in the same sample. As we have already seen, the
antiferromagnetic phase was suppressed completely at 150 K with an applied
magnetic field of 5 T.

There is a further change at about 190 K, where the magnetization unusually
decreases as a function of time. Again, it is very interesting to note that,
the temperature where this change in behavior is found is very close to the
one corresponding to the change from negative to positive LTLRR in the
relaxation procedure with {\em H}$_{end}${\em =0 T}. Although the change
with temperature of the LTLRR in figure 5c is very systematic, we must be
careful in this case because changes smaller than 1 \% are difficult to
separate from the experimental error (see scale in figure 5c).

The LTLRR temperature dependence observed in this last experiment with a
constant applied field ({\em H}$_{end}${\em =5 T}) is very similar (but with
inverted signs) to the previous case, when the applied field was removed,
leaving only a remanent magnetic field of about 1 mT. In our view, these
results eliminate the possibility that the residual field trapped in the
superconducting magnet of the SQUID or PPMS magnetometers could be the cause
for the observed unusual relaxations.

\subsection{Fitting of relaxation curves}

Our relaxation measurements at long time scales follow approximately a
logarithmic law: {\em M (t) / M (t}$_{n}${\em ) = 1 + S }$\cdot ${\em \ ln
(t / t}$_{n}${\em )}. Here, {\em S} is called magnetic viscosity, and{\em \ t%
}$_{n}$ and {\em M (t}$_{n}$) are the normalization time and the
corresponding magnetization at that moment, respectively\cite{Földeáki}$^{,}$
\cite{Chantrell}. This logarithmic relaxation has also been found in spin
glass systems\cite{Mydosh}, superconductor materials\cite{Campbell} and
mixture of small ferromagnetic particles\cite{Labarta}$^{,}$ \cite{Iglesias}.

The logarithmic relaxation has been attributed to the existence of a
distribution of energy barriers separating local minima in the free energy,
which correspond to different equilibrium states\cite{Földeáki}$^{-}$\cite
{Iglesias}. In our polycrystalline samples of La$_{0.5}$Ca$_{0.5}$MnO$_{3}$
and Nd$_{0.5}$Sr$_{0.5}$MnO$_{3}$ we have a spatially inhomogeneous mixture
of ferromagnetic and antiferromagnetic domains, which produce frustration in
the interactions among individual spins. This frustration was visualized
before in the differences among ZFC, FC and FCW curves in figure 1.

We performed the fitting of all the relaxation curves using the logarithmic
law mentioned above, where {\em S} was the only free\ parameter. In order to
get information over the whole temperature interval we consider {\em t}$_{n}$%
{\em \ = 1000 s}, removing from the fitting any transient behavior\ at the
beginning of each relaxation measurement. Due to the fact that at long times
the relaxation curves are linear in a semi-logarithmic plot, {\em S} could
be considered as a normalized value of the LTLRR.

Figure 6 shows the temperature dependence of {\em S} in the cases where the
external field is removed (H$_{end}$ = 0 T) for La$_{0.5}$Ca$_{0.5}$MnO$_{3}$
(close squares) and Nd$_{0.5}$Sr$_{0.5}$MnO$_{3}$ (open circles), and when
the external field is kept constant (H$_{end}$ = 5 T) for La$_{0.5}$Ca$_{0.5}
$MnO$_{3}$ (open up triangles). The corresponding error bars associated to
the fitting procedure are smaller than the symbols used in all cases and the
continuous lines are only guides to the eye. This figure displays the main
magnetic relaxation results in a single graph. It is interesting to note the
change in the temperature dependence of {\em S} between 130 K and 150 K.
Moreover, {\em S} changes sign between 180 and 195 K in all cases.

Note also that to facilitate the comparison among different data sets the
curve for Nd$_{0.5}$Sr$_{0.5}$MnO$_{3}$ was multiplied by 5 and the curve
for La$_{0.5}$Ca$_{0.5}$MnO$_{3}$ with H$_{end}$ = 5 T was multiplied by 20.
The decrease in the absolute value of {\em S} in Nd$_{0.5}$Sr$_{0.5}$MnO$_{3}
$\ in comparison with La$_{0.5}$Ca$_{0.5}$MnO$_{3}$ (H$_{end}$ = 0 T)\ was
expected due to the higher internal magnetic field experienced by Nd$_{0.5}$%
Sr$_{0.5}$MnO$_{3}$. As discussed before, this is associated with the
stronger ferromagnetic interactions in Nd$_{0.5}$Sr$_{0.5}$MnO$_{3}$. The
same effect, now due to the external magnetic field, was verified in the
comparison between the absolute values of {\em S} in La$_{0.5}$Ca$_{0.5}$MnO$%
_{3}$, when the measurements were done removing the field (H$_{end}$ = 0 T)\
and with the constant field (H$_{end}$ = 5 T).

Some reports\cite{Földeáki}$^{,}$ \cite{Chantrell} have claimed that as a
first approximation {\em S }could be considered proportional to ${\em (}$%
{\em k}$_{B}${\em $\cdot $T) / U}. In other words, the magnetic viscosity is
expected to have two competing factors: the thermal energy, which favors a
faster relaxation, and an effective pinning energy of the spin system, which
opposes to it. This could be viewed as an effective potential well where the
depth corresponds to {\em U} and the excitation energy to {\em k}$_{B}$$%
\cdot ${\em T}. However, it is important to stress here that this simple
model cannot explain the observed changes of sign in {\em S}.

If we consider the absolute value of {\em S} as the physically relevant
parameter, ignoring the changes in sign, we then find that below 150 K in La$%
_{0.5}$Ca$_{0.5}$MnO$_{3}$\ (130 K in Nd$_{0.5}$Sr$_{0.5}$MnO$_{3}$) the
thermal energy increases at a higher rate than {\em U} for increasing
temperatures. This is due to the increase in the absolute value of {\em S}
with temperature. On the other hand, between 150 K and 195 K in La$_{0.5}$Ca$%
_{0.5}$MnO$_{3}$\ (130 K and 180 K in Nd$_{0.5}$Sr$_{0.5}$MnO$_{3}$) the
absolute value of {\em S} decreases with rising temperatures. This result
suggests that the effective pinning energy grows in this interval at a
faster rate in comparison with the thermal energy. Above 195 K in La$_{0.5}$%
Ca$_{0.5}$MnO$_{3}$\ (180 K in Nd$_{0.5}$Sr$_{0.5}$MnO$_{3}$) the absolute
value of {\em S} increases again with temperature, and the predominant role
of the thermal energy is recovered.

We also tried to fit the relaxation curves to a stretched exponential
dependence: {\em M(t) = M(}$\infty ${\em ) - }$\left[ {\em M(0)\ -\ M(}%
\infty {\em )}\right] ${\em \ }$\cdot ${\em \ exp}$\left[ {\em -(t/\ }\tau 
{\em )}^{{\em n}}\right] ${\em \ }, where {\em M(0)} and {\em M(}$\infty $%
{\em )} are the magnetizations at times {\em t = 0} and{\em \ t = }$\infty $%
, $\tau $ is a characteristic relaxation time and {\em n} is a parameter
that could change between 0 and 1. The value of {\em n=1} corresponds to a
single exponential dependence, characteristic of only one energy barrier in
the free energy. However, if {\em 0 
\mbox{$<$}%
n 
\mbox{$<$}%
1}, that would mean that a distribution of energy barriers and relaxation
times are present in the system. This expression has been used before for
spin glass systems \cite{Mydosh} and also for La$_{0.5}$Ca$_{0.5}$MnO$_{3}$ 
\cite{Smolyaninova}.

The main problem with this kind of fitting is that manganese perovskite
samples have a characteristic relaxation time much bigger than that of the
usually available total measuring time. This makes the estimation of {\em M(}%
$\infty ${\em )} very difficult. For example, Smolyaninova et. al. \cite
{Smolyaninova}, after 24 hours measuring the magnetic relaxation in a
similar sample of La$_{0.5}$Ca$_{0.5}$MnO$_{3}$ at 12 K, did not find
saturation in the magnetization. Trying to solve this problem we
approximated $\left[ {\em M(t)-\ M(}\infty {\em )}\right] ${\em \ / }$\left[ 
{\em M(0)-\ M(}\infty {\em )}\right] $ by {\em M(t) / M(0)}. Even though the
fittings are not very good for all temperatures, we found as an order of
magnitude for La$_{0.5}$Ca$_{0.5}$MnO$_{3}$ at 150 K and {\em H}$_{end}${\em %
\ = 0 T}, values of $\tau ${\em =10}$^{7}${\em \ s} (approximately 100 days)
and {\em n=0.3}. Repeating the same procedure at 130 K for Nd$_{0.5}$Sr$%
_{0.5}$MnO$_{3}$\ we obtained values of $\tau ${\em =10}$^{8}${\em \ s}
(about 3 years) and {\em n=0.3}. These results reinforce the idea of the
long relaxation times and the wide distribution of energy barriers involved
in these samples.

\section{Conclusions}

We performed a systematic study of magnetic relaxation curves after applying
and removing a 5 T magnetic field, in polycrystalline samples of La$_{0.5}$Ca%
$_{0.5}$MnO$_{3}$ and Nd$_{0.5}$Sr$_{0.5}$MnO$_{3}$. The LTLRR in La$_{0.5}$%
Ca$_{0.5}$MnO$_{3}$ ({\em H}$_{end}${\em \ = 0 T}) decreased from 10 to 150
K and increased from 150 to 195 K. This change in behavior was found to be
correlated with the complete destruction of the antiferromagnetic phase in
the presence of a 5 T magnetic field above 150 K. At 195 K, the
magnetization decreased initially in a very short time interval and after
that it increased slowly as a function of time. Moreover, between 200 and
245 K, an increase of magnetization above {\em M(0)}, was observed. The
change from a negative slope to a positive one at about 200 K was found to
be related to the suppression of antiferromagnetic fluctuations with small
magnetic fields.

A similar temperature dependence of the LTLRR was found for the Nd$_{0.5}$Sr$%
_{0.5}$MnO$_{3}$ sample. However, the temperature where the LTLRR reached
the minimum in Nd$_{0.5}$Sr$_{0.5}$MnO$_{3}$ was lower than in La$_{0.5}$Ca$%
_{0.5}$MnO$_{3}$, in agreement with the stronger ferromagnetic interactions
in Nd$_{0.5}$Sr$_{0.5}$MnO$_{3}$. This suggested that the general
temperature dependence of the LTLRR, and the corresponding physics, were
mainly independent of the particular charge-ordering sample considered.

We have also measured the relaxation curves in the La$_{0.5}$Ca$_{0.5}$MnO$%
_{3}$ sample with a constant magnetic field of 5 T. The LTLRR values in this
case showed a temperature dependence similar to the previous ones, but with
inverted signs. They increased from 10 to 150 K and decreased from 150 to
210 K. The peak in the temperature dependence of the LTLRR was again around
150 K. These measurements also eliminated doubts about a possible influence
of the small magnetic field trapped in the superconductor solenoid of the
SQUID magnetometer after removing the field in the standard procedure.

We successfully performed the fitting of all the relaxation curves using a
logarithmic law. The slow relaxation was attributed to the coexistence of
ferromagnetic and antiferromagnetic interactions, which produced a
distribution of energy barriers. The decrease of the absolute value of {\em S%
} with rising temperatures between 150 and 195 K in La$_{0.5}$Ca$_{0.5}$MnO$%
_{3}$\ (130 and 180 K in Nd$_{0.5}$Sr$_{0.5}$MnO$_{3}$) suggested that,
contrary to the other intervals, here the effective pinning energy grew at a
faster rate than the thermal energy. Besides, a stretched exponential
dependence of the relaxation curves at 150 K for La$_{0.5}$Ca$_{0.5}$MnO$_{3}
$, and at 130 K for\ Nd$_{0.5}$Sr$_{0.5}$MnO$_{3}$, showed the existence of
very long relaxation times and a wide distribution of pinning energies.

Although further studies would be required to fully understand the
temperature dependence of the relaxation in charge ordering compounds, our
preliminary findings indicate a correlation between the instability of two
competing magnetic phases and the unusual magnetic relaxation. These
intriguing results in the relaxation measurements could be a consequence of
the competition between ferromagnetic double exchange and antiferromagnetic
superexchange interactions.

{\Large Acknowledgments}

We thank the Brazilian Science Agencies FAPESP, CAPES, CNPq and PRONEX for
financial support. We also acknowledge Prof. A. V. Narlikar for a careful
revision of the text.

\bigskip

\bigskip {\em Figure Captions}

Figure 1. Magnetization vs. temperature, measured with\ an applied magnetic
field of 5 T for (a) La$_{0.5}$Ca$_{0.5}$MnO$_{3}$ and (b) Nd$_{0.5}$Sr$%
_{0.5}$MnO$_{3}$\ samples. The insets show the same type of measurements
with (a) H=1.2 mT and (b) H=0.1 mT. Magnetization is given in Bohr magnetons
per manganese ion. Arrows show the direction of temperature sweep. The large
hysteresis makes these compounds particularly interesting to study their
relaxation behavior.

Figure 2. Magnetization vs. applied magnetic field for representative
temperatures in La$_{0.5}$Ca$_{0.5}$MnO$_{3}$ (a and b) and Nd$_{0.5}$Sr$%
_{0.5}$MnO$_{3}$ (c and d) samples. Magnetization is given in Bohr magnetons
per manganese ion. The applied magnetic field was cycled in all cases from 0
to 5 T and then back to 0 T again. Ferromagnetic and antiferromagnetic
interactions coexist in both compounds, leading to charge and orbital
ordered phases.

Figure 3. Normalized magnetic relaxation measurements, after applying and
removing a magnetic field of 5 T, in a La$_{0.5}$Ca$_{0.5}$MnO$_{3}$ sample:
from (a) 10 to 150 K, (b) 150 to 195 K and (c) 195 to 245 K. Time is shown
in logarithmic scale. The diagram in figure 3a represents the evolution in
time of the applied field. The insets in figure 3b and 3c reproduce details
of the curve at 195 and 200 K prior to normalization and with the
corresponding error bars. In the last case the error bars have the same
dimension of the circles.

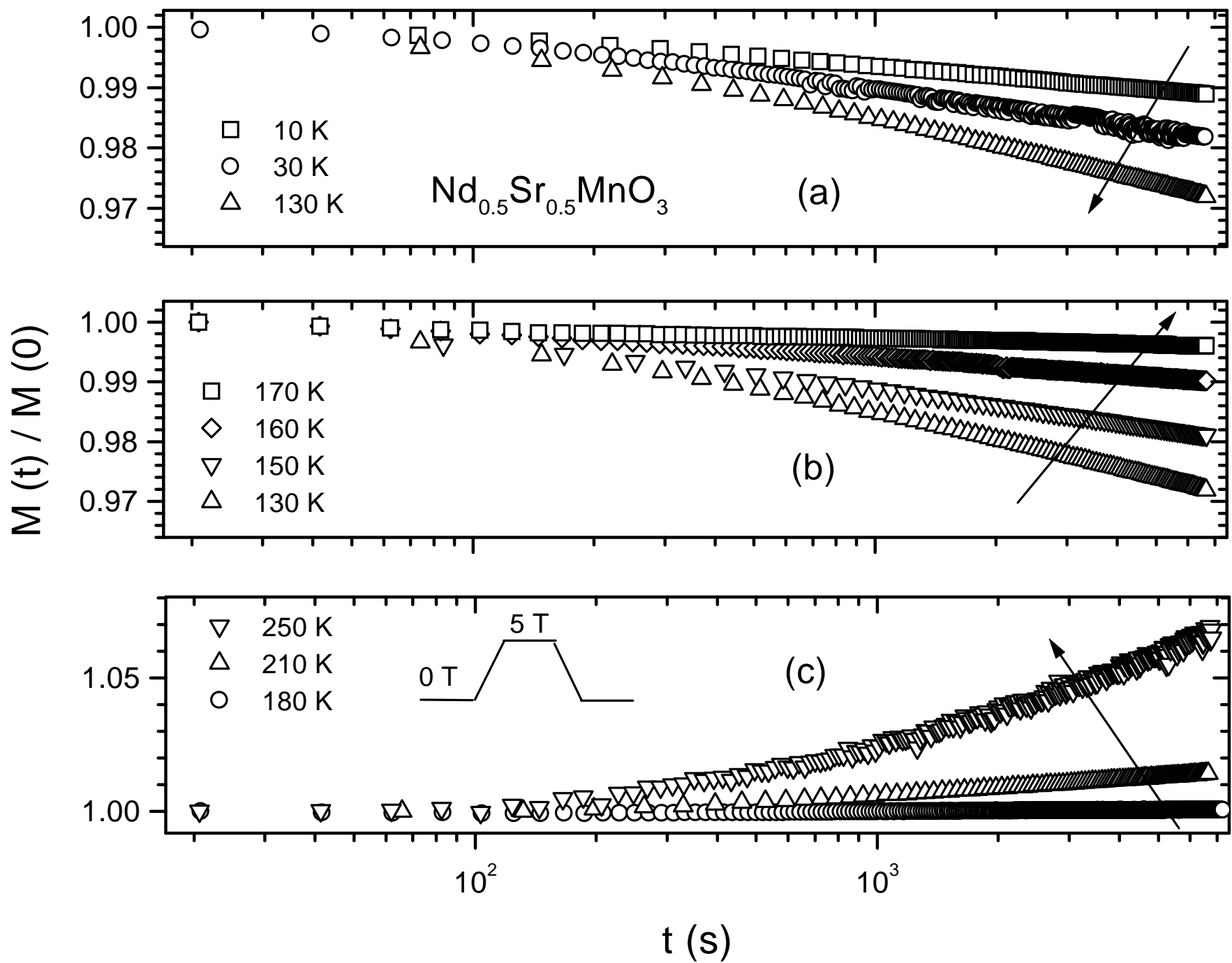
Figure 4. Normalized magnetic relaxation measurements, after applying and
removing a magnetic field of 5 T, for a Nd$_{0.5}$Sr$_{0.5}$MnO$_{3}$
sample. The diagram represents the evolution in time of the applied magnetic
field. The curves show the same qualitative behavior found in the La$_{0.5}$%
Ca$_{0.5}$MnO$_{3}$ compound.

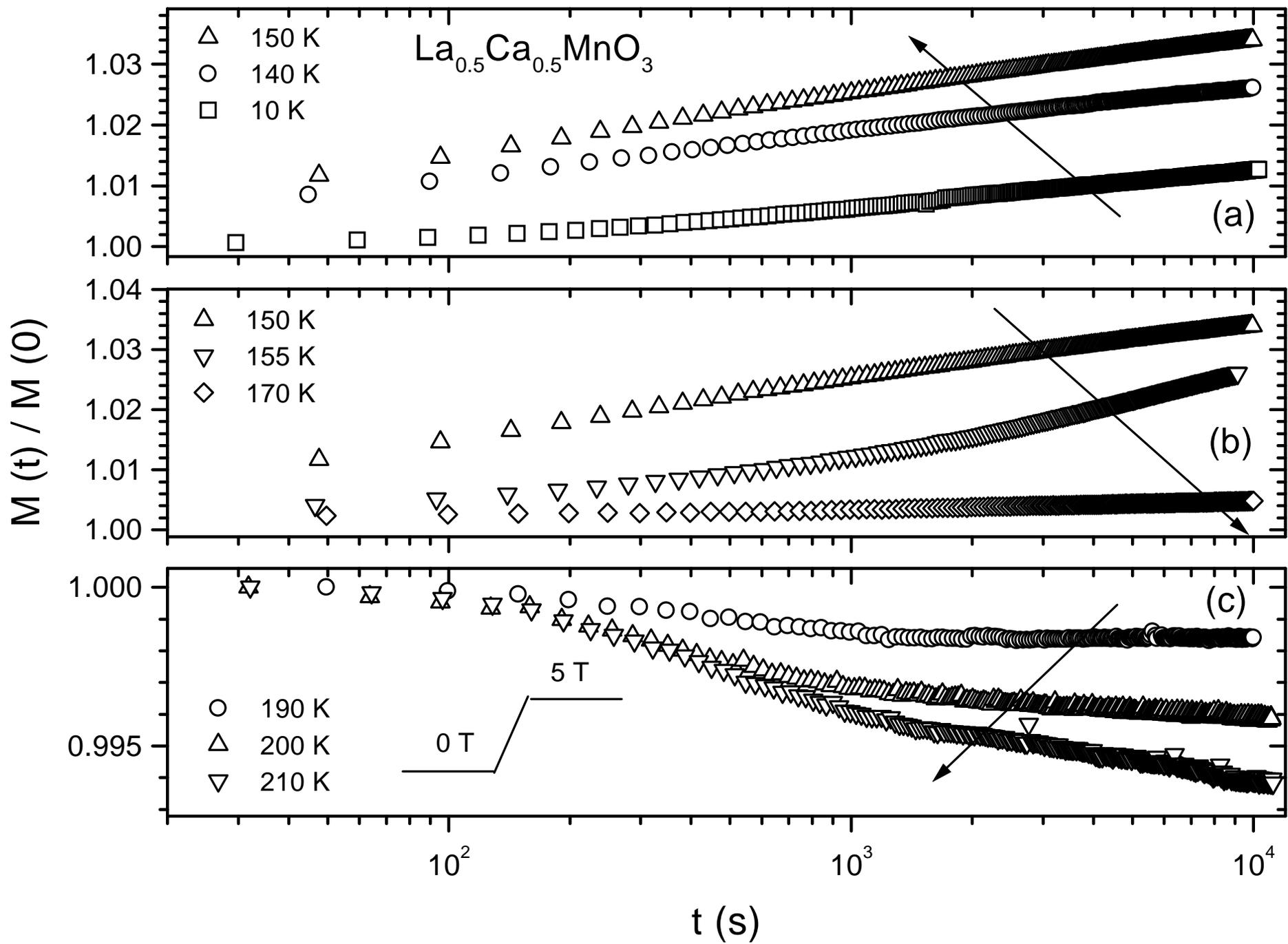
Figure 5. Normalized magnetic relaxation measurements in the presence of 5 T
magnetic field for a La$_{0.5}$Ca$_{0.5}$MnO$_{3}$ sample. The diagram
represents the evolution in time of the applied field. The temperature
dependence of the relaxation rate values with {\em H}$_{end}${\em =5 T} are
similar (but with inverted signs) to the ones with {\em H}$_{end}${\em =0}.

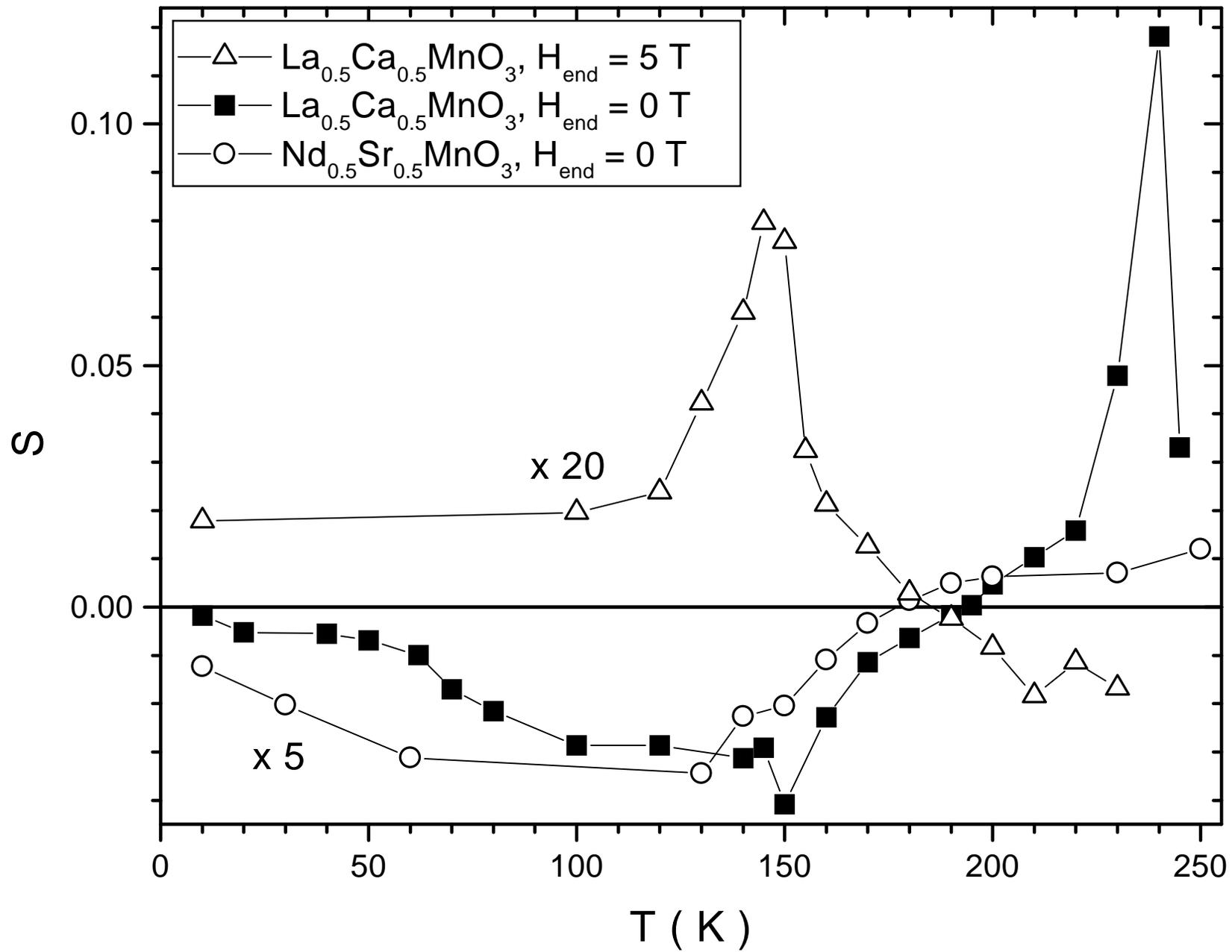
Figure 6. Results of the fitting of the curves in figures 3, 4 and 5 with a
logarithmic law at long time scales. {\em S} is the magnetic viscosity,
which is the only fitting parameter. Measurements were done removing the
final external field (H$_{end}$ = 0 T) for La$_{0.5}$Ca$_{0.5}$MnO$_{3}$
(closed squares) and Nd$_{0.5}$Sr$_{0.5}$MnO$_{3}$ (open circles), and with
the final external field constant (H$_{end}$ = 5 T) for La$_{0.5}$Ca$_{0.5}$%
MnO$_{3}$ (open up triangles). Lines are only guides to the eye. Note that
to facilitate the comparison among different data sets the curve for Nd$%
_{0.5}$Sr$_{0.5}$MnO$_{3}$ was multiplied by 5 and the curve for La$_{0.5}$Ca%
$_{0.5}$MnO$_{3}$ with H$_{end}$ = 5 T was multiplied by 20.

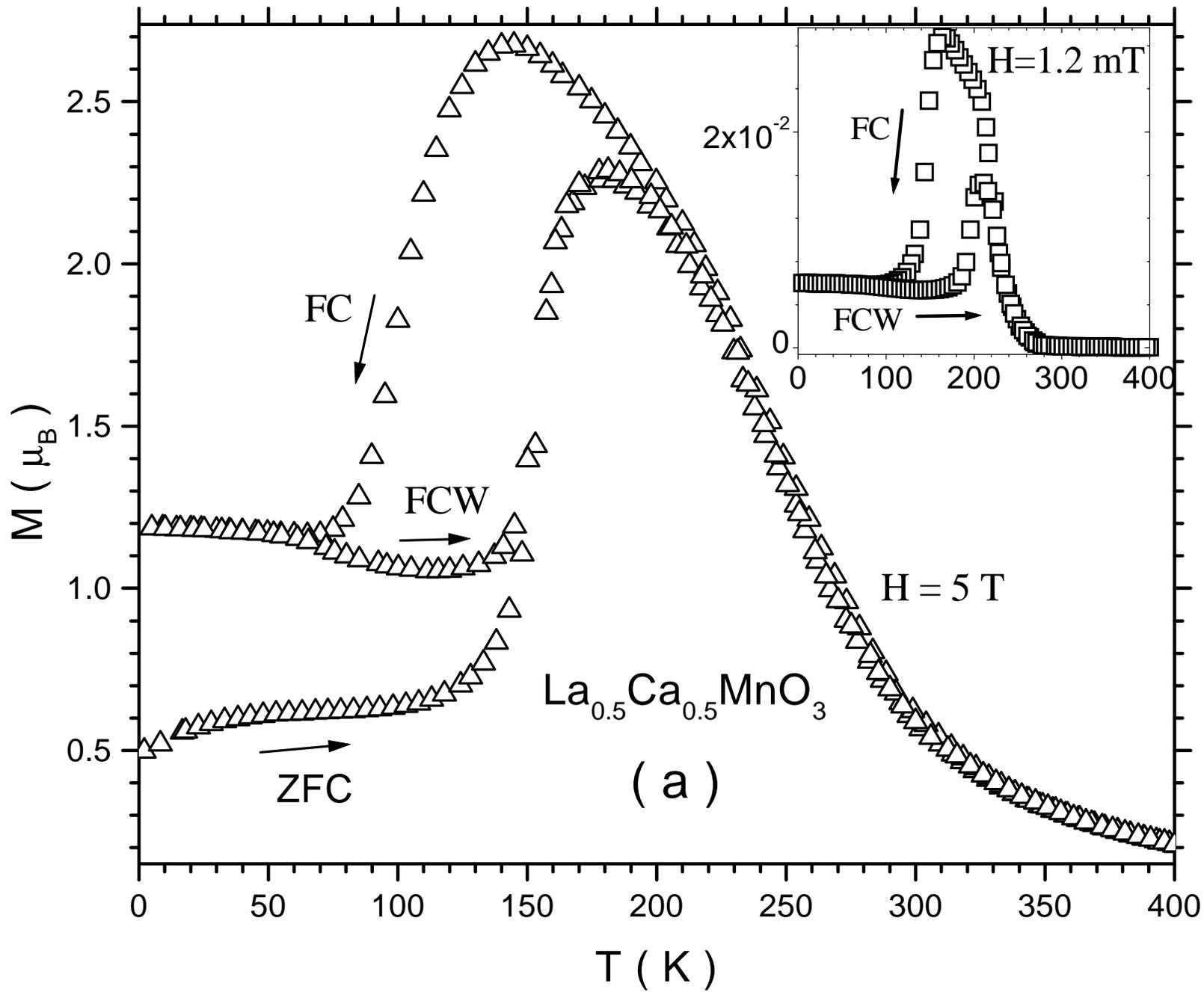
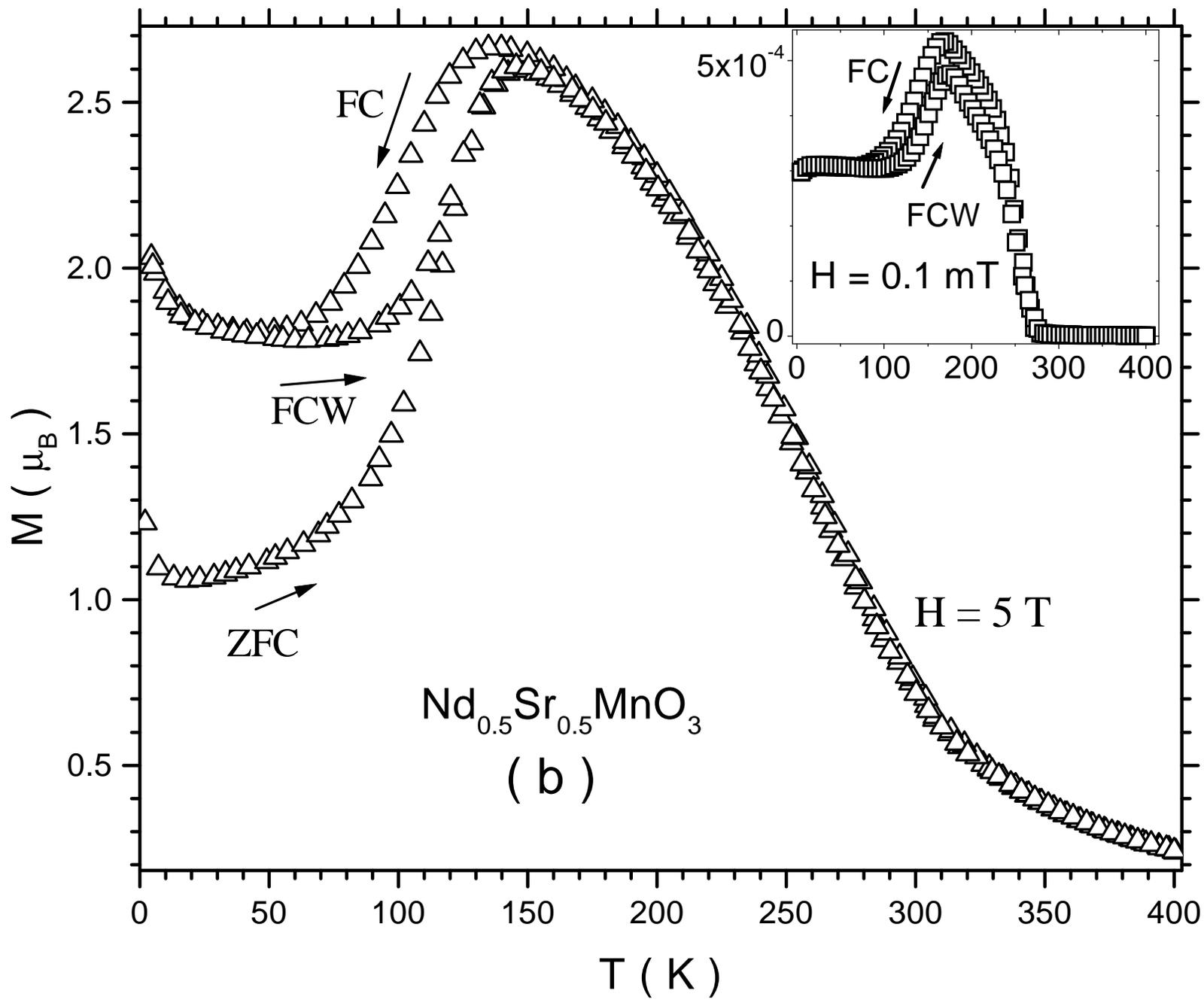
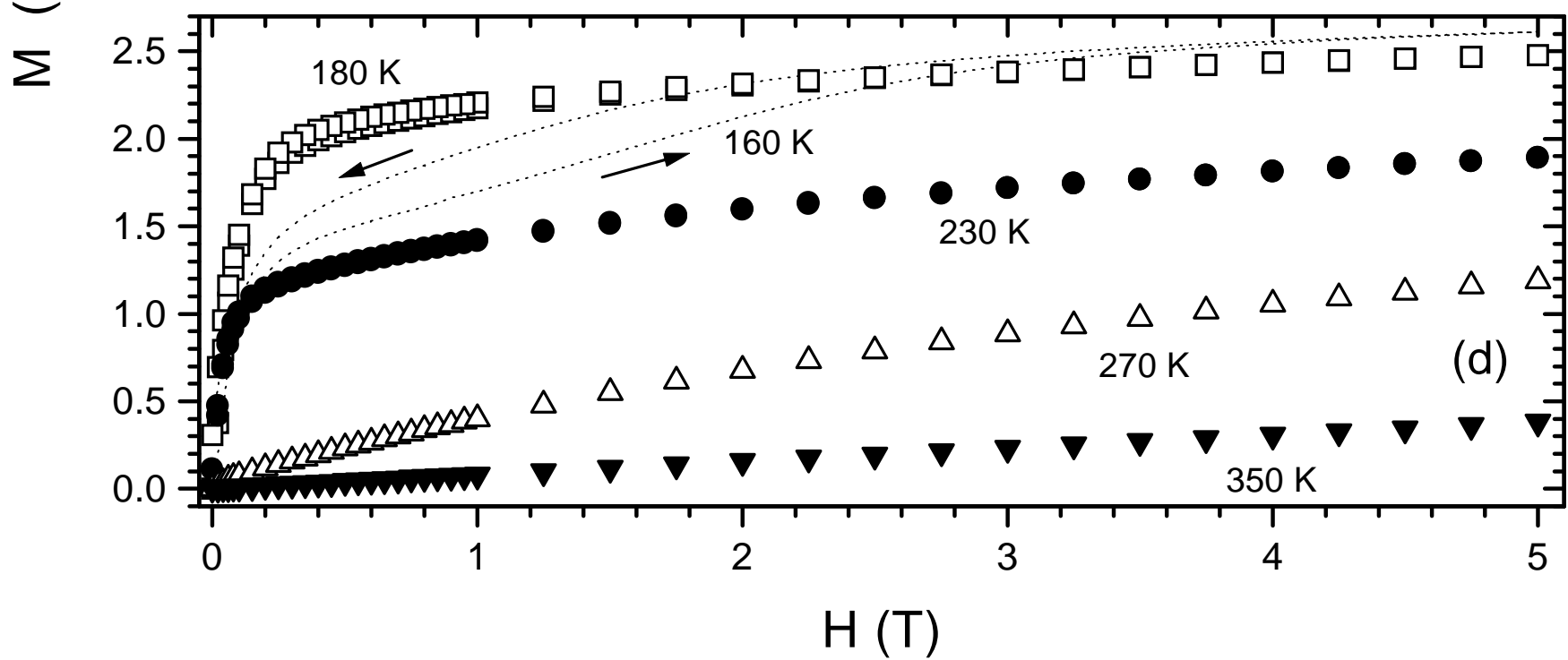
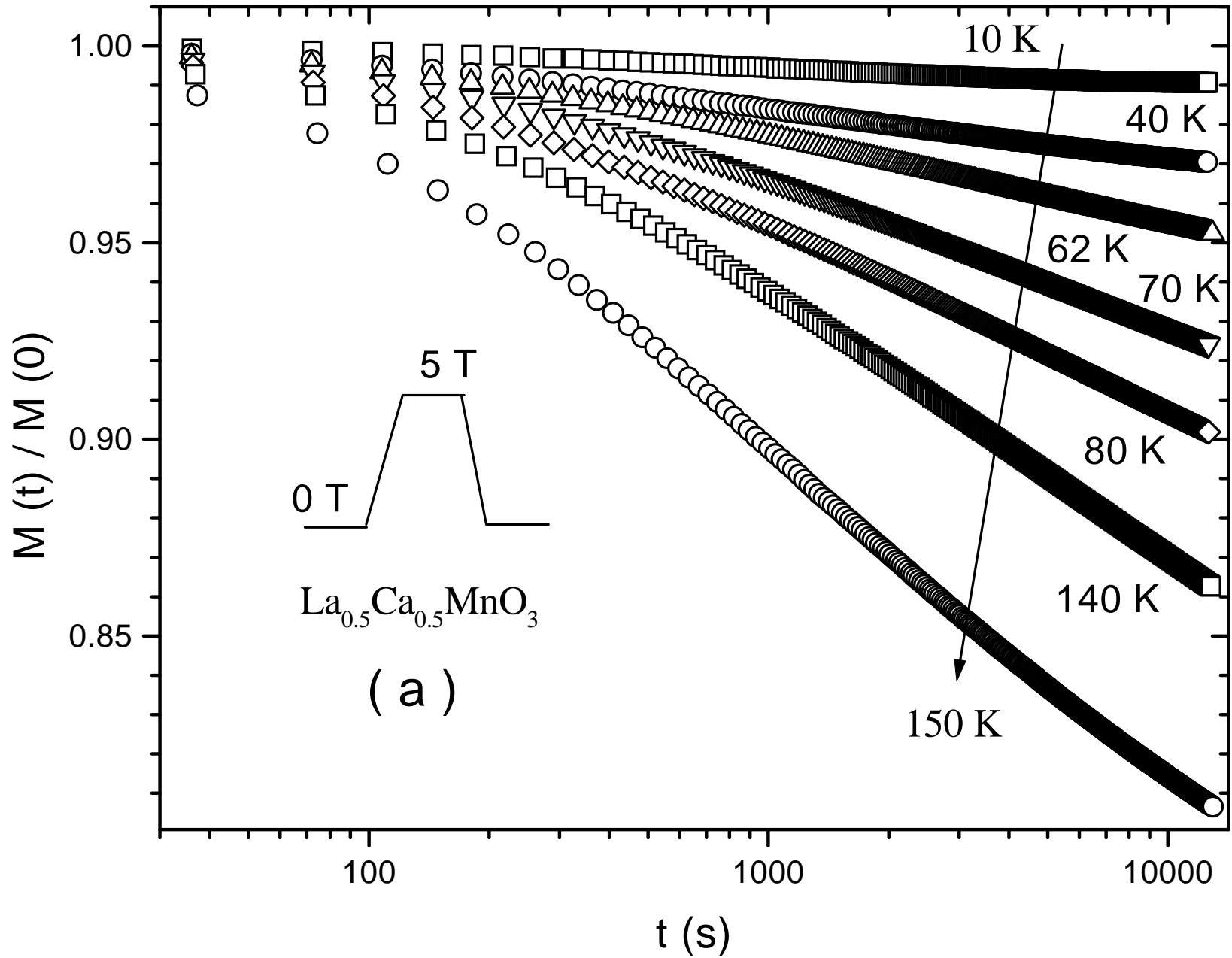
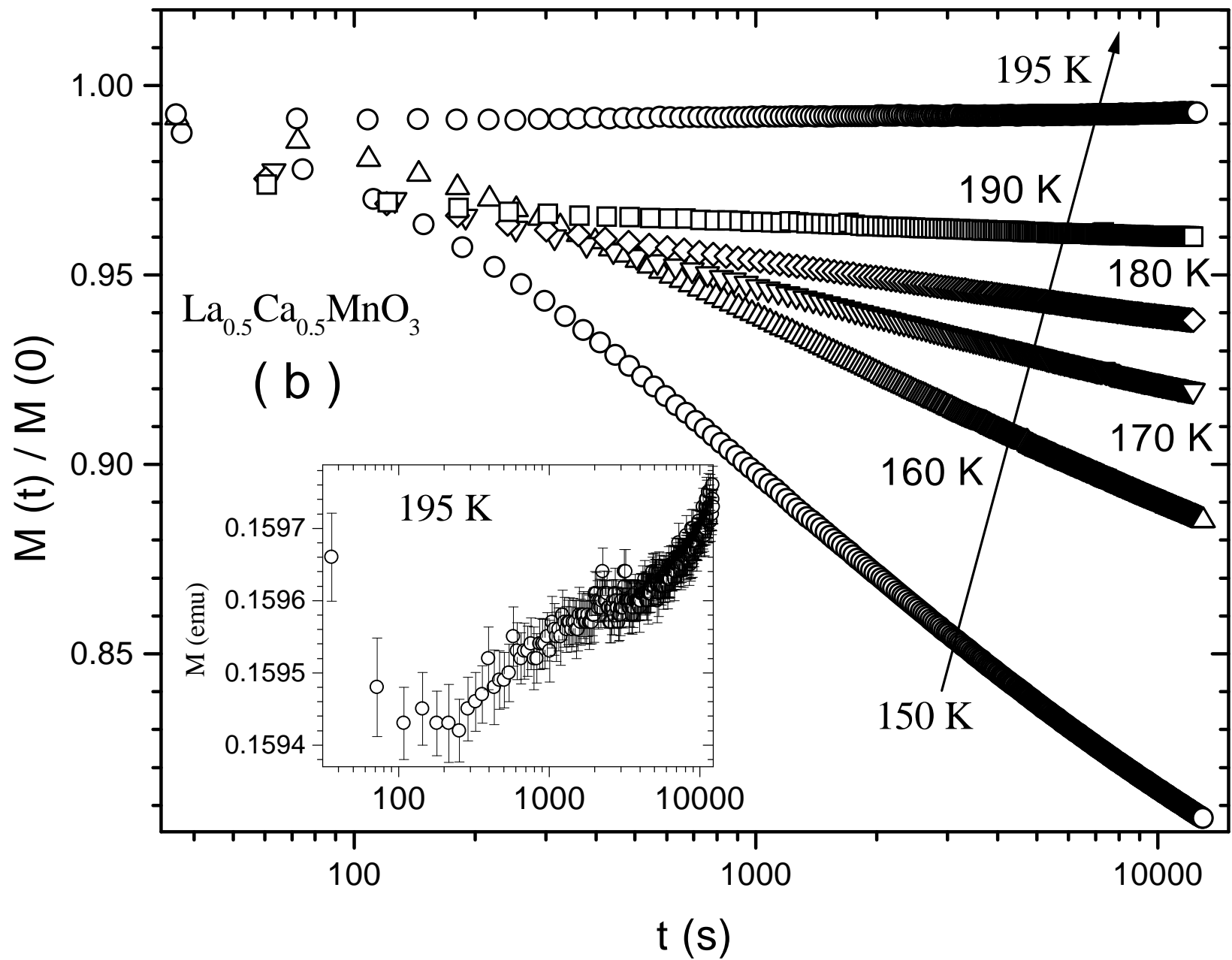
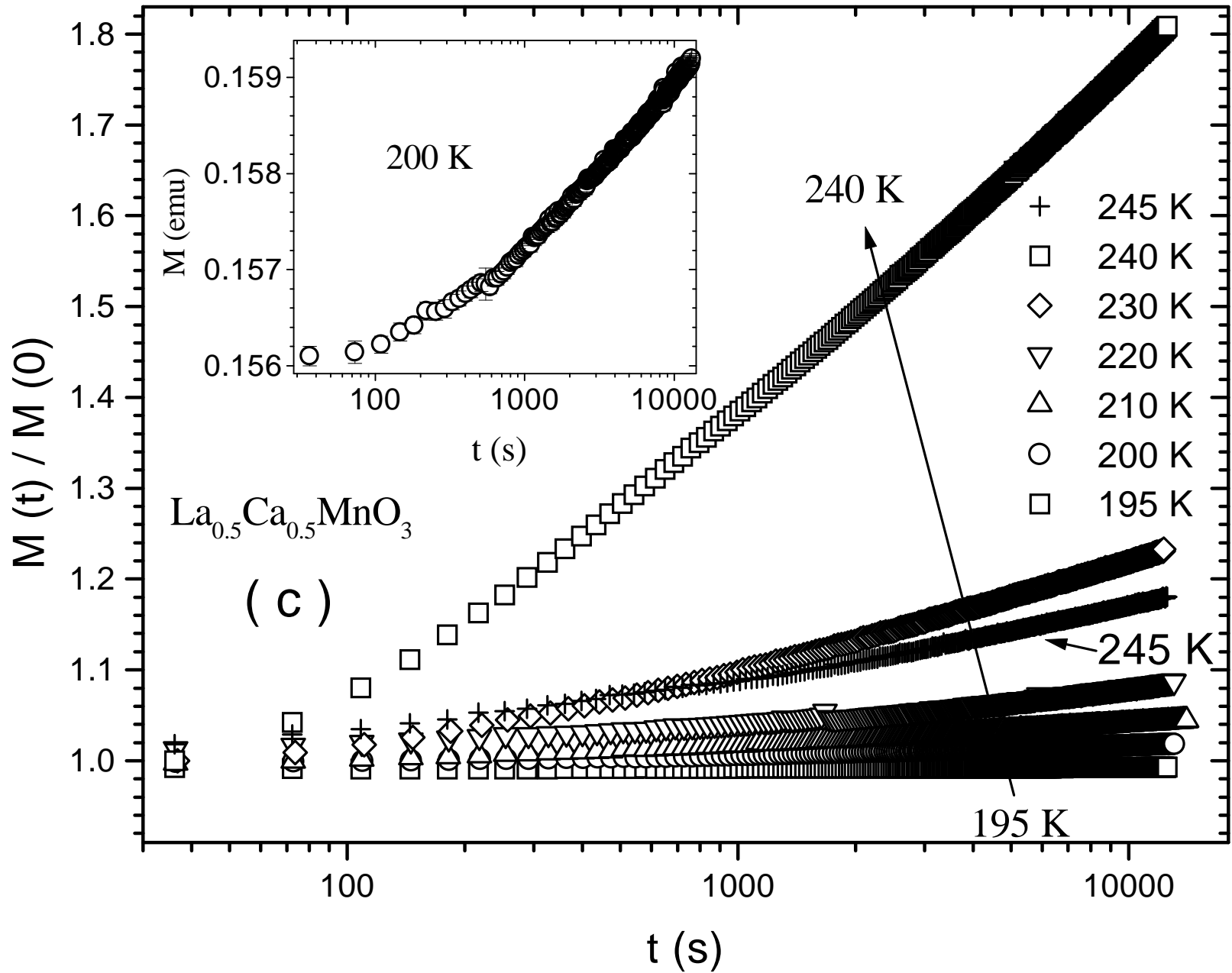
\end{document}